\documentclass{ws-procs9x6}

\newcommand{\slsh}[1]{\mbox{$\not\! #1$}}

\begin{document}

\title{Quark and Gluon Sivers Functions\footnote{\uppercase{T}his work is
supported by \uppercase{F}ondecyt (\uppercase{C}hile) under grant
1030355.}}

\author{Ivan Schmidt
}

\address{Departamento de F\'\i sica,
Universidad T\'ecnica Federico Santa Mar\'\i a, \\ Casilla 110-V,
Valpara\'\i so, Chile \\ E-mail: ischmidt@fis.utfsm.cl}

%

\maketitle

\abstracts{ The physics of hadron single transverse spin
asymmetries is discussed. Possible measurements of both the quark
and gluon Sivers functions are proposed.}

\section{Introduction}

In the usual QCD factorization formalism, a collinear
approximation for the partonic intrinsic motion is used, and
therefore the total inclusive hadronic cross section is written as
a convolution of a hard elementary partonic cross section with
distribution and fragmentation functions in which the transverse
motion of the partons has been integrated. Nevertheless, the
intrinsic quark and gluon transverse momenta are important,
because for one thing they provide corrections to the collinear
approximation, and moreover they are essential in order to explain
single spin asymmetries (SSA) within a generalized transverse
momentum dependent QCD factorization formalism. In this case the
cross section for an inclusive process $AB \to CX$ is written as:

\begin{eqnarray}
d\sigma & = &\sum\limits_{abc} f_{a/A} (x_a ,k_{ \bot a} ) \otimes
f_{b/B} (x_b ,k_{ \bot b} ) \label{eq:murn}\\ &&\otimes
d\mathord{\buildrel{\lower3pt\hbox{$\scriptscriptstyle\frown$}}
\over \sigma } ^{ab \to c...} (x_a ,x_b ,k_{ \bot a} ,k_{ \bot b}
) \otimes D_{C/c} (z,k_{ \bot C} )\nonumber
\end{eqnarray}

In dealing with SSAs, the most important transverse momentum
dependent functions are the Sivers distribution function $f_{1T}^
\bot = f_{a/p \uparrow } (x_a ,\vec k_{ \bot a} ) - f_{a/p
\downarrow } (x_a ,\vec k_{ \bot a} )$, which gives the
probability distribution of finding unpolarized quarks inside a
transversely polarized proton, and the Collins fragmentation
function $H_1^ \bot$, which gives the probability of unpolarized
hadrons coming from the fragmentation of a transversely polarized
quark. In several processes both functions can contribute.

\section{Quark Sivers Function}

The Sivers function is proportional to the T-odd correlation $\vec
S_\bot \cdot  (\vec P \times \vec k_ \bot  )$, and therefore
contains an azimuthal asymmetry with respect to the direction of
the hadron momentum $\vec P$. In 1993 Collins gave a proof about
the vanishing of this function\cite{Collins1993}, but which was
shown later on by Collins himself to be incorrect.

In order to understand the physics that is present in the Sivers
function let us consider a specific process: semi-inclusive deep
inelastic lepton scattering $\ell p^\uparrow \to \ell^\prime \pi
X$. In the target rest frame, single-spin correlations correspond
to the $T$-odd triple product $i \vec S_p \cdot \vec p_\pi \times
\vec q,$ where the phase $i$ is required by time-reversal
invariance.  The differential cross section thus has an azimuthal
asymmetry proportional to $|\vec p_{\pi}||\vec q| {\rm sin}
\theta_{q \pi} {\rm sin} \phi$ where $\phi$ is the angle between
the plane containing the photon and pion and the plane containing
the photon and proton polarization vector $\vec S_p.$ In a general
frame, the azimuthal asymmetry has the invariant form ${i\over M}
\epsilon_{\mu \nu \sigma \tau} P^\mu S_p^\nu p^\sigma_\pi q^\tau $
where the polarization four-vector of the proton satisfies $S_p^2
= -1$ and $S_p \cdot P = 0.$

In general we can express the SSA as:
\begin{equation}
SSA=\frac{{\left\langle  \uparrow  \right|\left.  \uparrow
\right\rangle  - \left\langle  \downarrow  \right|\left.
\downarrow  \right\rangle }}{{\left\langle  \uparrow \right|\left.
\uparrow  \right\rangle  + \left\langle  \downarrow \right|\left.
\downarrow  \right\rangle }},
\end{equation}
where the transverse spin  basis is related to the helicity basis
by $ \left| { \uparrow / \downarrow } \right\rangle  =
\frac{1}{{\sqrt 2 }}\left( {\left|  +  \right\rangle  \pm i\left|
- \right\rangle } \right)$, which means that the SSA can be
written as:
\begin{equation}
\frac{{2{\mathop{\rm Im}\nolimits} \left\langle + \right|\left. -
\right\rangle }}{{\left\langle  + \right|\left. + \right\rangle  +
\left\langle  -  \right|\left.  - \right\rangle }}
\end{equation}
Therefore in order to produce a correlation involving a
transversely-polarized proton, there are two necessary conditions:
(1) There must be two proton spin amplitudes $M[{\gamma^*
p(J^z_p)\to F}]$ with $J^z_p = \pm {1\over 2}$ which couple to the
same final-state $|F>$; and (2) The two amplitudes must have
different, complex phases. The correlation is proportional to
${\rm Im}(M[J^z_p=+{1\over 2}]^*M[J^z_p=-{1\over 2}])$.

As a result, we can reach the following conclusions: (1) The
analysis of single-spin asymmetries requires an understanding of
QCD at the amplitude level.  (2) It also provides a handle on the
proton angular momentum. Since we need the interference of two
amplitudes which have different proton spin $J^z_p= \pm {1\over
2}$ but couple to the same final-state, the orbital angular
momentum of the two proton wavefunctions must differ by $\Delta
L^z = 1.$ The anomalous magnetic moment for the proton is also
proportional to the interference of amplitudes $M[{\gamma^*
p(J^z_p) \to F}]$ with $J^z_p = \pm {1\over 2}$ which couple to
the same final-state $|F>$. (3) Since we need an Imaginary part,
the SSA cannot come from tree level diagrams.

Final state interactions clearly fit into this picture. If a
target is stable, its light-front wavefunction must be real.  Thus
the only source of a nonzero complex phase in leptoproduction in
the light-front frame are final-state interactions. The
rescattering corrections from final-state exchange of gauge
particles produce Coulomb-like complex phases which, however,
depend on the proton spin.

In Ref. 2 the single-spin asymmetry in semi-inclusive
electroproduction $\gamma^* p \to H X$, induced by final-state
interactions, was calculated in a model of a spin-1/2 proton of
mass $M$ with charged spin-1/2 and spin-0 constituents of mass $m$
and $\lambda$, respectively, as in the QCD-motivated quark-diquark
model of a nucleon. The basic electroproduction reaction is then
$\gamma^* p \to q (qq)_0$.

There it was shown that the final-state interactions from gluon
exchange between the outgoing quark and the target spectator
system leads to single-spin asymmetries in deep inelastic
lepton-proton scattering at leading twist in perturbative QCD;
{\em i.e.}, the rescattering corrections are not power-law
suppressed at large photon virtuality $Q^2$ at fixed $x_{bj}$. The
azimuthal single-spin asymmetry  transverse to the photon-to-pion
production plane decreases as $ \alpha_s(r^2_\perp) x_{bj} M
r_\perp [\ln r^2_\perp]/ {r}_{\perp}^2$ for large $r_\perp,$ where
$r_\perp$ is the magnitude of the momentum of the current quark
jet relative to the virtual photon direction. The fall-off in
$r^2_\perp$ instead of  $Q^2$  compensates for the dimension of
the $\bar q$-$q$ -gluon correlation. The mass $M$ of the physical
proton mass appears here since it determines the ratio of the $L_z
= 1$ and $L_z = 0$ matrix elements. This is the same type of
physics that gives shadowing and antishadowing effects in both
electromagnetic and weak deep inelastic scattering in
nuclei\cite{BSY}.

A related analysis also predicts that the initial-state
interactions from gluon exchange between the incoming quark and
the target spectator system lead to leading-twist single-spin
asymmetries in the Drell-Yan process $H_1 H_2^\updownarrow \to
\ell^+ \ell^- X$\cite{Collins,BHS2}. These final- and
initial-state interactions can be identified as the path-ordered
exponentials which are required by gauge invariance and which
augment the basic light-front wavefunctions of
hadrons\cite{Collins,Ji:2002aa}. Both pictures, final and initial
state interactions and different gauge links, lead to the
conclusion that the Sivers function is not really universal, but
changes sign between SSAs in deep inelastic scattering and Drell-
Yan processes.

\section{How to obtain the Sivers function ?}

There have been many theoretical and experimental analysis about
ways in which to separate the Collins and Sivers effects. Probably
the simplest is to use the SSAs which can be measured in weak
interaction processes. For example, consider charged current
neutrino semi-inclusive deep inelastic scattering, where a hadron
(pion) is measured in the final state. In this case, the
transversity distribution cannot contribute to the cross section
since the produced quark from the weak interaction of the $W$
boson is always left-handed. On the other hand, in the final-state
interaction picture the SSA in charged and neutral current weak
interactions will also be present, just as in the electromagnetic
case.  Thus these weak interaction processes will clearly
distinguish the underlying physical mechanisms which produce
target single-spin asymmetries\cite{BHS3}.

Let us see this in more detail. The quark distribution in the
proton is described by a correlation matrix:
\begin{equation}
\Phi^{\alpha\beta} (x,{\bm p}_{\perp}) = \int {d\xi^-d^2{\bm
\xi}_{\perp}\over (2\pi)^3}e^{ip\cdot \xi}
<P,S|{\bar{\psi}}^{\beta}(0)\psi^{\alpha}(\xi)|P,S>\mid_{\xi^+=0}\
, \label{w1}
\end{equation}
where $x=p^+/P^+$. The correlation matrix $\Phi$ is parameterized
in terms of the transverse momentum dependent quark distribution
functions\cite{MB98}:
\begin{eqnarray}
\Phi (x , {\bm p}_{\perp}) &=& {1\over 2}{\Bigg[}f_1{\slsh{n}}+
f_{1T}^{\perp} {{\epsilon}_{\mu\nu\rho\sigma}{\gamma}^{\mu}
n^{\nu}p_{\perp}^{\rho}S_{\perp}^{\sigma}\over M}
+g_{1s}{\gamma}_5{\slsh{n}} \label{w2}\\ &&+h_{1T}
i{\gamma}_5{\sigma}_{\mu\nu}n^{\mu}S_{\perp}^{\nu} +h_{1s}^{\perp}
{i{\gamma}_5{\sigma}_{\mu\nu}n^{\mu}p_{\perp}^{\nu}\over M}
+h_1^{\perp} {{\sigma}_{\mu\nu}p_{\perp}^{\mu}n^{\nu}\over M}
{\Bigg]}\ , \nonumber
\end{eqnarray}
where the distribution functions have arguments $x$ and ${\bm
p}_{\perp}$, and $n^\mu =(n^+,n^-,{\bm n}_{\perp}) =(0,2,{\bm
0}_{\perp})$.

Similar expressions and parametrization can be obtained for the
quark fragmentation correlation matrix $\Delta^{\alpha\beta} (z ,
{\bm k}_{\perp})$.

\subsection{Electromagnetic case}
The hadronic tensor of the leptoproduction by the electromagnetic
interaction in leading order in $1/Q$ is given by\cite{MB98}
\begin{eqnarray}
2M{ W}^{\mu\nu}(q,P,P_h)&=& \int d^2{\bm p}_{\perp} d^2{\bm
k}_{\perp} \delta^2({\bm p}_{\perp}+{\bm q}_{\perp}-{\bm
k}_{\perp}) \nonumber\\ &&\times {1\over 4}\ {\rm Tr}\Big[ \Phi
(x_B,{\bm p}_{\perp})\gamma^\mu \Delta (z_h,{\bm
k}_{\perp})\gamma^\nu \Big] \nonumber\\ &&+\ \Bigl(\
q\leftrightarrow -q\ ,\ \ \mu \leftrightarrow \nu\ \Bigr)\ ,
\label{w5}
\end{eqnarray}
where $x_B=Q^2/2P\cdot q$ and $z_h=P\cdot P_h/P\cdot q$. The
momentum ${\bm q}_{\perp}$ is the transverse momentum of the
exchanged photon in the frame where $P$ and $P_h$ do not have
transverse momenta.

The single-spin asymmetry (SSA) in semi-inclusive deep inelastic
scattering (SIDIS) $ep^{\updownarrow}\to e'\pi X$, which is given
by the correlation ${\vec S}_p\cdot {\vec q}\times {\vec
p}_{\pi}$, is obtained from (\ref{w5}).  As mentioned before, for
the electromagnetic interaction there are two mechanisms for this
SSA: $h_1 H_1^{\perp}$ and $f_{1T}^{\perp}D_1$ (Collins and Sivers
effects), where $h_1$ is the transversity distribution and $D_1$
the unpolarized quark fragmentation function. .

We can also consider the SSA of $e^+e^-$ annihilation processes
such as $e^+e^-\to \gamma^* \to \pi {\Lambda}^{\updownarrow} X$.
The $\Lambda$ reveals its polarization via its decay $\Lambda \to
p \pi^-$.  The spin of the $\Lambda$ is normal to the decay plane.
Thus we can look for a SSA through the T-odd correlation
$\epsilon_{\mu \nu \rho \sigma} S^\mu_\Lambda p^\nu_\Lambda
q^\rho_{\gamma^*} p^\sigma_{\pi}$.  This is related by crossing to
SIDIS on a $\Lambda$ target.

\subsection{Charged weak current case}
{\bf Charged currents:}~  Let us consider the SSA in the charged
current (CC) weak interaction process $\nu p^{\updownarrow}\to
\ell\pi X$. For the CC weak interaction, the trace in (\ref{w5})
becomes
\begin{equation}
{\rm Tr}\Big[ \Phi \gamma^\mu P_L \Delta \gamma^\nu P_L \Big]
=
{\rm Tr}\Big[ \Phi P_R\gamma^\mu P_L \Delta P_R \gamma^\nu P_L
\Big]
=
{\rm Tr}\Big[ \Phi_{\rm CC} \gamma^\mu \Delta_{\rm CC} \gamma^\nu
\Big] \ , \label{w6}
\end{equation}
where $P_L=(1-\gamma_5)/2$, $P_R=(1+\gamma_5)/2$, and
\begin{equation}
\Phi_{\rm CC}\equiv P_L \Phi P_R\ ,\qquad \Delta_{\rm CC}\equiv
P_L \Delta P_R\ . \label{w7}
\end{equation}

The result is that $\Phi_{\rm CC}$ does not contain the chiral-odd
distribution functions which are present in (\ref{w2}), and
$\Delta_{\rm CC}$ does not contain the corresponding chiral-odd
fragmentation functions. The charged current only couples to a
single quark chirality, and thus it is not sensitive to the
transversity distribution. Thus  SSAs can only arise in  charged
current weak interaction SIDIS from the Sivers FSI mechanism
$f_{1T}^{\perp}D_1$ in leading order in $1/Q$; in contrast, both
the Collins  $h_1 H_1^{\perp}$ and Sivers $f_{1T}^{\perp}D_1$
mechanisms contribute to SSAs for the electromagnetic and neutral
current (NC) weak interactions.

In an similar way, we can also consider the SSAs of the processes
$\pi p^{\updownarrow}\ ({\rm or}\ p p^{\updownarrow})\to W X\to
\ell \nu X.$ If $y$ denotes the $W$ rapidity, it can be shown that
the region $y\sim -1$ is very sensitive to the antiquark Sivers
functions, whereas the region $y\sim +1$ is sensitive to the quark
Sivers functions. Furthermore, it turns out that the SSA for $W^+$
gives information about the  Sivers $u$ quark distribution in the
region $y \to 1$ and about the Sivers $\bar d$ in the region $y
\to -1$. Something similar happens for the $W^-$ SSA
(interchanging $u$ and $d$). Therefore the measurement of the SSAs
$A_N^{W^{\pm}}$ is a practical way to separate the $u$ and $d$
quarks Sivers functions and their corresponding antiquark
distributions $\bar u$ and $\bar d$~\cite{SS03}.

{\bf Neutral currents:}~Let us now consider the SSA in the neutral
current weak interaction process $\nu p^{\updownarrow}\to \nu\pi
X$.  For the NC weak interaction, the interaction vertex of
$Z$-f-f is given by $(-ie/{\rm sin}{\theta}_{\rm W}{\rm
cos}{\theta}_{\rm W}) (c_LP_L+c_RP_R)$ with the weak
isospin-dependent coefficients $c_{L,R}=I^3_{\rm W}-Q{\rm
sin}^2{\theta}_{\rm W}$.  Explicit values of $c_{L,R}$ are given
by $c_L={1\over 2}-{2\over 3}{\rm sin}^2{\theta}_{\rm W}$,
$c_R=-{2\over 3}{\rm sin}^2{\theta}_{\rm W}$ for u, c, t quarks,
and $c_L=-{1\over 2}+{1\over 3}{\rm sin}^2{\theta}_{\rm W}$,
$c_R={1\over 3}{\rm sin}^2{\theta}_{\rm W}$ for d, s, b quarks.

The trace in (\ref{w5}) becomes
\begin{eqnarray}
&&a\ {\rm Tr}\Big[ \Phi \gamma^\mu (c_LP_L+c_RP_R) \Delta
\gamma^\nu (c_LP_L+c_RP_R) \Big] \label{nc1}\\ &=& a\ {\rm
Tr}\Big[ \Phi (c_LP_R+c_RP_L) \gamma^\mu \Delta \gamma^\nu
(c_LP_L+c_RP_R) \Big] \ =\ a\ {\rm Tr}\Big[ \Phi_{\rm NC}
\gamma^\mu \Delta \gamma^\nu \Big] \ , \nonumber
\end{eqnarray}
where $a=1/{\rm sin}^2{\theta}_{\rm W}{\rm cos}^2{\theta}_{\rm W}$
and
\begin{equation}
\Phi_{\rm NC}\equiv (c_LP_L+c_RP_R)\Phi (c_LP_R+c_RP_L)\ .
\label{nc2}
\end{equation}

In this case, for the Sivers effect we find that the SSA is given
by that of the electromagnetic case with $f_{1T}^{\perp}D_1$
replaced by
\begin{equation}
a\ {c_L^2 + c_R^2\over 2}\ f_{1T}^{\perp}D_1\ . \label{nc12}
\end{equation}
However, $f_1$ is also weighted by the same factor $a\, (c_L^2 +
c_R^2) / 2$. Therefore, the SSA from the final-state interaction
mechanism in the NC weak interaction is the same as that in the
electromagnetic interaction. This can be confirmed in the simple
quark-diquark model.

For the $h_1 H_1^{\perp}$ mechanism, we find that the SSA is given
by that of the electromagnetic case with $(h_1 H_1^{\perp})/(f_1
D_1)$ replaced by
\begin{equation}
{2c_Lc_R \over c_L^2 + c_R^2}\ {h_1 H_1^{\perp} \over f_1 D_1} \ .
\label{nc11}
\end{equation}
That is, the SSAs are modified by the quark weak isospin-dependent
factor $2c_Lc_R / (c_L^2 + c_R^2)$ in comparison with the
electromagnetic case. The same factor appears in the linear $\cos
\theta$ forward-backward asymmetry in the $ e^+ e^- \to Z \to q
\bar q$ reaction.

The SSA of the Drell-Yan processes at the  $Z^0$, such as $\pi
p^{\updownarrow}\ ({\rm or}\ p p^{\updownarrow})\to Z X\to
\ell^+\ell^- X$, can arise from the $h_1 h_1^{\perp}$ and
$f_{1T}^{\perp}D_1$ mechanisms. We can also consider the SSA of
the $e^+e^-$ annihilation processes such as $e^+e^-\to Z \to \pi
{\Lambda}^{\updownarrow} X$, which can arise from the $H_1
H_1^{\perp}$ and $D_{1T}^{\perp}D_1$ mechanisms\cite{Boer:1997qn}.
The SSAs of these processes have the same situation as those of
the above SIDIS case. The initial/final-state interaction
mechanisms have the same formulas as the electromagnetic case,
whereas the Collins mechanisms are weighted by the quark weak
isospin-dependent factor $2c_Lc_R / (c_L^2 + c_R^2)$ present in
(\ref{nc11}).

\section{Gluon Sivers function}

The gluon Sivers function was mentioned for the first time in Ref.
9, and recently it was also considered in jet
correlations\cite{BV} and in $D$ meson production\cite{ABALM} in
$p^{\uparrow}p$ collisions.

The direct photon production in $pp$ collisions can provide a
clear test of short-distance dynamics as predicted by perturbative
QCD, because the photon originates in the hard scattering
subprocess and does not fragment, which immediately means that the
Collins effect is {\it not} present\cite{SSY}. This process is
very sensitive to the gluon structure function, since it is
dominated by the quark-gluon Compton subprocess in a large photon
transverse momentum range. Prompt-photon production, $pp
(p\bar{p}) \to \gamma X$, has been a useful tool for the
determination of the unpolarized gluon density and it is
considered one of the most reliable reactions for extracting
information on the polarization of the gluon in the
nucleon\cite{BSSV}.

It turns out that both Sivers functions for quarks and gluons are
involved in the SSA for direct photon production
$A^{\gamma}(s,x_F)$, and therefore it is necessary to identify a
kinematic region where the gluon Sivers function dominates. The
cross section contains two terms, the first one involves a product
of the quark Sivers function at the light-cone momentum variable
$x_b$ and the transverse momentum dependent unpolarized gluon
distribution at light-cone momentum $x_a$, while the second
involves a product of the gluon Sivers function at $x_b$ times the
transverse momentum dependent unpolarized quark distribution at
$x_a$.  Thus it is necessary to determine the range of integration
over $x_a$ and to study the relative magnitude of $x_a$ and $x_b$.
As an example, taking  $\sqrt{s}=200~\rm{GeV}$ and
$p_T=20~\rm{GeV}$, we find that the minimum value of $x_a$,
$x_{min}\approx x_F$ in the region $x_F > 0.3 $. On the other
hand, we can also see that when $x_a$ is integrated over the range
[$x_{min}$, 1], the main contribution comes from the low $x_b$
values. Therefore, when we look at the large $x_F$ region, where
$x_a$ is large but $x_b$ is small, the asymmetry can be
approximately expressed as
\begin{equation}
A^{\gamma}(s,x_F) = \frac{ \langle \Delta_N G \rangle } { \langle
G \rangle }~,
\end{equation}
where $\langle \Delta_N G \rangle$ and $\langle G \rangle$ mean
the corresponding values over an appropriate integrating range.

\section*{Acknowledgments}
The work presented here was done in collaboration with Stanley
Brodsky, Dae Sung Hwang, and Jacques Soffer.

\end{document}